\documentclass[12pt,a4paper]{article}
\usepackage[utf8]{inputenc}
\usepackage[affil -it]{authblk}
\usepackage{graphicx}
\usepackage{verbatim}
\usepackage{amsmath,amsfonts,amssymb,textcomp}
\usepackage{color}
\usepackage{authblk}

\begin{document}

\title{The role of electron scattering from registration detector in  a MAC-E type spectrometer}
\author[1,2]{P.\,V.\,Grigorieva}
\author[1,2]{A.\,A.~Nozik}
\author[1]{V.\,S.\,Pantuev}
\author[1]{A.\,K.~Skasyrskaya}

\affil[1]{Institute for Nuclear Research of Russian Academy of Sciences, Moscow, Russia}
\affil[2]{Moscow Institute of Physics and Technology, Dolgoprudny, Russia}

\maketitle

\begin{abstract}
There is a proposal to search for a sterile neutrino in a few keV mass range by the "Troitsk nu-mass" facility. In order to estimate sterile neutrino mixing one needs to make precision spectrum measurements well below the endpoint using the existing electrostatic spectrometer with a magnetic adiabatic collimation, or MAC-E filter. The expected signature will be a kink in the electron energy spectrum in tritium beta-decay. In this article we consider the systematic effect of electron backscattering on the detector used in the spectrometer. For this purpose we provide a set of Monte-Carlo simulation results of electron backscattering on a silicon detector with a thin golden window with realistic electric and magnetic fields in the spectrometer. We have found that the probability of such an effect reaches up to 20-30\%. The scattered electron could be reflected backwards to the detector by electrostatic field or by magnetic mirror. There is also a few percent probability to escape from the spectrometer through  its entrance. A time delay between the scattering moment on the detector and the return of the reflected electron can reach a couple of microseconds in the Troitsk spectrometer. Such estimations are critical for the planning upgrades of the detector and the registration electronics.

\end{abstract}

%\linenumbers

\section{Introduction}
The "Troitsk nu-mass" program is conducted by the Institute for Nuclear Research of the Russian Academy of Sciences. The original measurement to set the limit on mass of electron anti-neutrino by analyzing the tritium beta-decay  spectrum was completed and the final results of these efforts were published in~\cite{Aseev:2011dq}. Currently we are expanding the energy range of measurements up to 5 keV from the endpoint with a goal to probe sterile neutrinos with masses up to 4 keV~\cite{our_nus}. 

At the first stage of the new program the "Troitsk nu-mass" experiment will utilize the same layout as in the past measurements~\cite{Aseev:2011dq} but with a new spectrometer. 
Compared to the old one, the spectrometer has a vessel with a diameter twice as big, volume  ten times larger and magnetic field of the main magnet stronger by about 20\%. There are some options of what kind of detector and readout electronics are needed to meet our requirements. Before making decision, we simulate response of the whole spectrometer system concerning tritium beta-spectrum measurement far from its end point. In this work we present the details of our spectrometer, then describe steps and results of electron scattering calculation and finally make conclusions. We have to mention that all our results described here are relevant to future experiments  using a MAC-E filter spectrometer, like  the KATRIN ~\cite{katrin} extended program on search for a sterile neutrino, or PTOLEMY experiment with the goal to detect relic neutrinos~\cite{ptolemy}.  

\section{Troitsk nu-mass MAC-E filter}
\begin{figure}[ht]
\centering
   \includegraphics[width=0.9\linewidth]{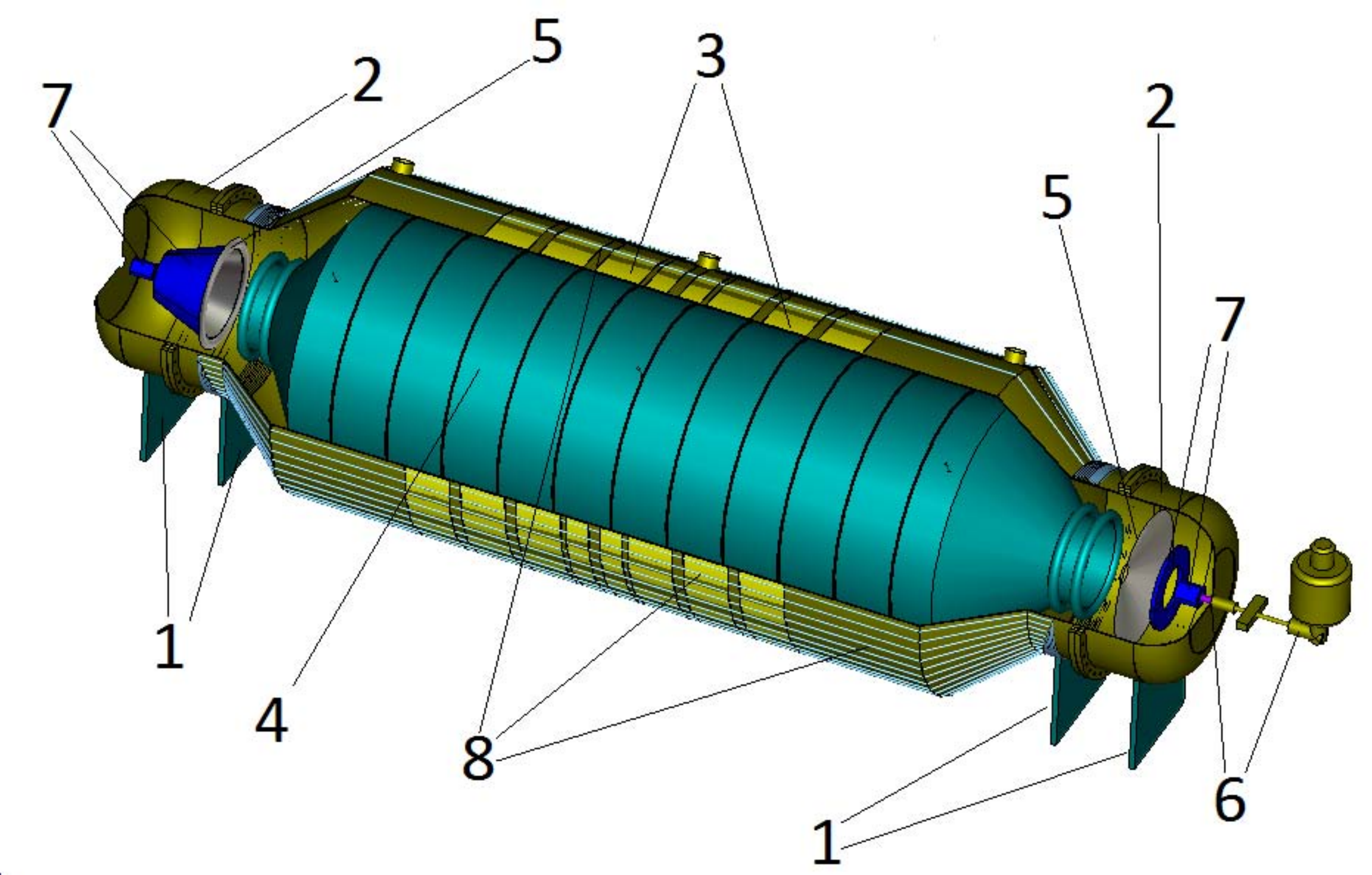}
   \caption{General view of the spectrometer. 1-  supports, 2 – side cups, 3 – axial winding, 4 – main high voltage electrode, 5 – additional ground electrodes, 6 – detector with liquid $N_2$ dewar, 7 - superconducting solenoids, 8 –  correction coils }
	\label{fig:spectrometer}  
\end{figure}

The spectrometer itself consists of a central vessel and two removable side caps,  Fig.~\ref{fig:spectrometer}. The vessel is a stainless steel cylinder with a conical taper at its edges for providing transition from the diameter of the cup to the diameter of the housing central part. The Sectional high voltage electrode is kept under common high voltage and is mounted on insulators. 
%%%%%%%%%%

\begin{figure}[ht]
\centering
   \includegraphics[width=0.7\linewidth]{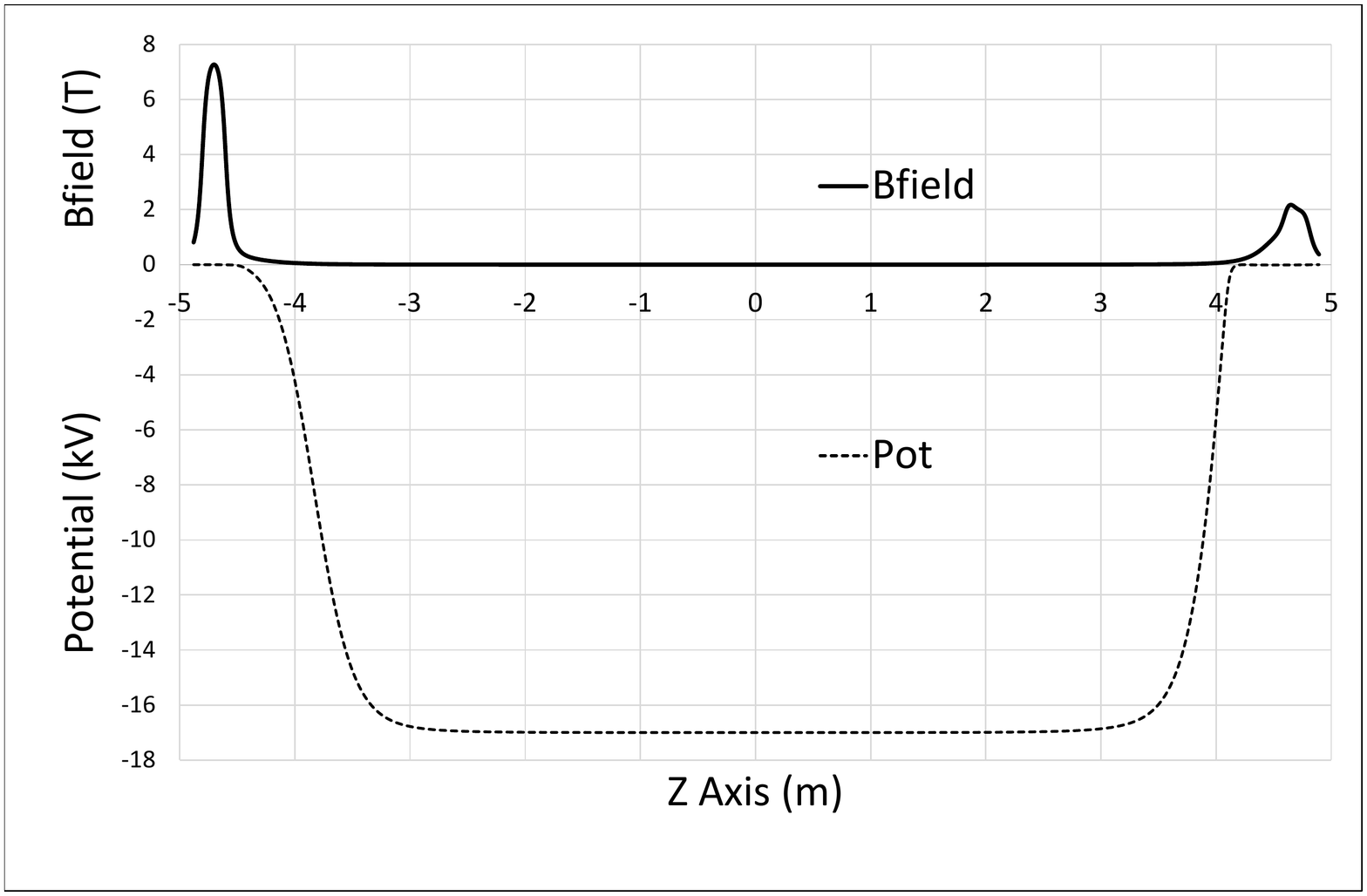}
   \caption{The strength of magnetic and electrostatic fileds in the spectrometer at its Z-axis. Zero Z is in the center of the spectrometer, the entrance -- on the left, the detector -- on the right. }
	\label{fig:fields}  
\end{figure}

The energy spectrum of $\beta$-electrons is measured in the spectrometer by varying the electrostatic potential applied. The electrons with energy smaller than this potential are reflected. The electrons with higher energy pass the electrostatic barrier and hit the detector. This is the way how the integral spectrum of electrons is measured. However, the retarding potential changes only the longitudinal component of the energy, $ E_{||} \propto v_{||}^2$. Therefore, in order to get good a energy resolution one has to make the transversal energy of electrons, $ E_{\perp} \propto v_{\perp}^2$, as small as possible inside the potential barrier, i.e. in the analyzing plane of the spectrometer.
This is achieved by a special configuration of electric and magnetic fields, Fig.~\ref{fig:fields}.

 The magnetic field lines of the spectrometer form a ``bottle'' shape with the highest longitudinal field at the pinch-magnet located near the spectrometer entrance (on the left in Fig.~\ref{fig:spectrometer} and Fig.~\ref{fig:fields}). The field in the pinch-magnet, $B_0$, is up to 8 Tesla. The lowest longitudinal magnetic field is in the analyzing plane in the center of the  spectrometer (Fig.~\ref{fig:fields}). The magnetic moment of a charged particle, $\mu=E_{\perp}/2B$, is an adiabatic invariant when the transversal gradient of the magnetic field is small. In this case we have $ E_{\perp m} = E_{\perp 0} \cdot B_m/B_0$,  where subscripts $m$ and $0$ refer to quantities in the analyzing mid-plane and in the pinch-magnet correspondingly. Therefore, the resolution of the spectrometer (spread of  transversal energies of electrons) is $ \Delta E _{\perp m} = E_{0} \cdot B_m/B_0$, which boils down to $\approx$ 1.5~eV at the highest energies, $E_0 \approx 18$ keV.  More details and an accurate formula are presented in \cite{our_nus}. The overall magnetic and electrostatic fields form the so called electrostatic spectrometer with a magnetic adiabatic collimation or MAC-E filter. Electrons from beta-decay in the gaseous source are transported by the solenoid field into the spectrometer entrance with a high magnetic field formed by the pinch magnet. Electrons move to the center of the spectrometer following the magnetic lines and reach there a very low magnetic field of the order of 6 Gs. Particles which energy is high enough to pass the retarding electrostatic potential are then collected by another magnet and detected by a surface barrier Si detector.
%%%%%%%%%%

The magnetic field in the spectrometer is formed by a system of two main electromagnets. Superconducting cryogenic magnets produce a field of up to 8 T at the input of the spectrometer (pinch-magnet) and up to 3 T at the detection side. The magnets are rather small, so the field quickly decreases with distance from the coils  and in the center of the spectrometer reaches  3 Gs. Fine tuning of the field to the desired shape in the central part of the spectrometer is controlled by warm electromagnetic coils wound outside the spectrometer vessel.  These four coils generate an additional axial field up to 4 Gs in the central analyzing plane of the spectrometer. The Earth's magnetic field and other external transversal fields are compensated by two warm  coils which can form 1.2 Gs field in the transverse vertical and horizontal directions. 

The additional calculations have shown that adiabatic approximation for the magnetic moment $\mu=E_{\perp}/2B$ for 18.5 keV electrons is valid down to 13.5 kV on the high voltage electrode, which covers current measurement region. A question remains what the electron behavior  will be if it scatters backwards on the detector and there is a big difference between the electron energy and the spectrometer electrostatic potential? To answer this question we have to consider the scattering first.

\section{Detector and electron scattering}
The current detector is  a surface barrier Si(Li), 25 mm in diameter, with a gold plated 20 $\mu \rm g/cm^2$ or 10 nm thick entrance window. The entrance window defines  about 2~keV threshold. The detector aperture is limited by a copper collimator, 17 mm in diameter. There are a few options to replace this detector and electronics  but we use this one as a typical example because backward electron scattering is a common feature for any kind of detectors. To simulate electron passage through the detector window and silicon we used an open Monte Carlo program CASINO~\cite{casino}. It is designed to simulate a large amount of electron trajectories in a solid. An important feature of the CASINO code is that it allows one to calculate  electrons with energies in a keV region.

\begin{figure}[htb]
\begin{minipage}[t]{0.48\linewidth}
\includegraphics[width=1.\textwidth]{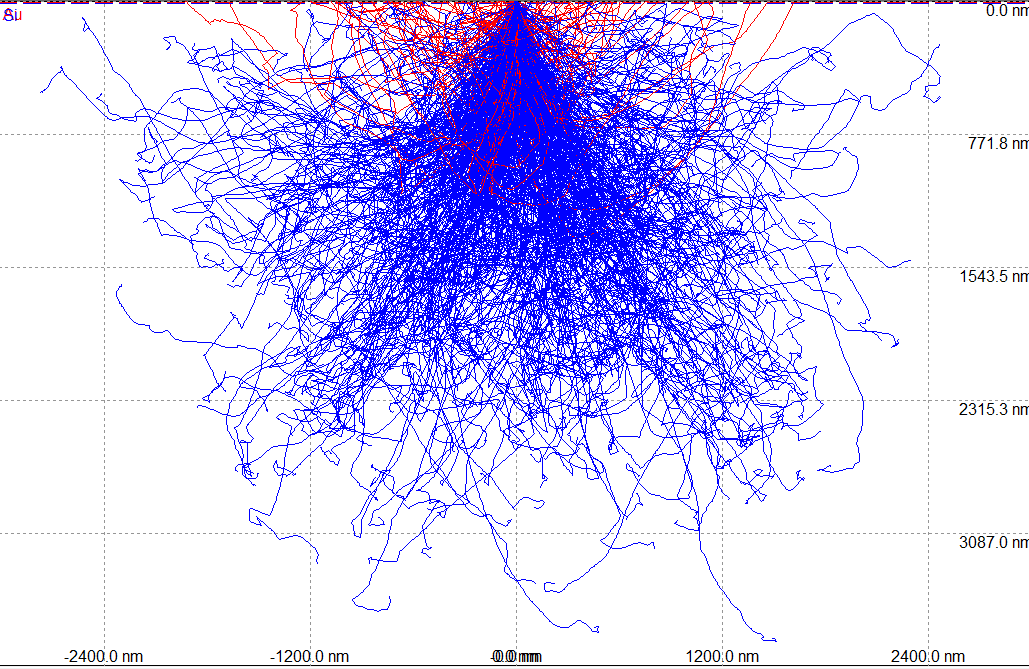}
\end{minipage}
\hfill
\begin{minipage}[t]{0.48\linewidth}
\includegraphics[width=1.08\textwidth]{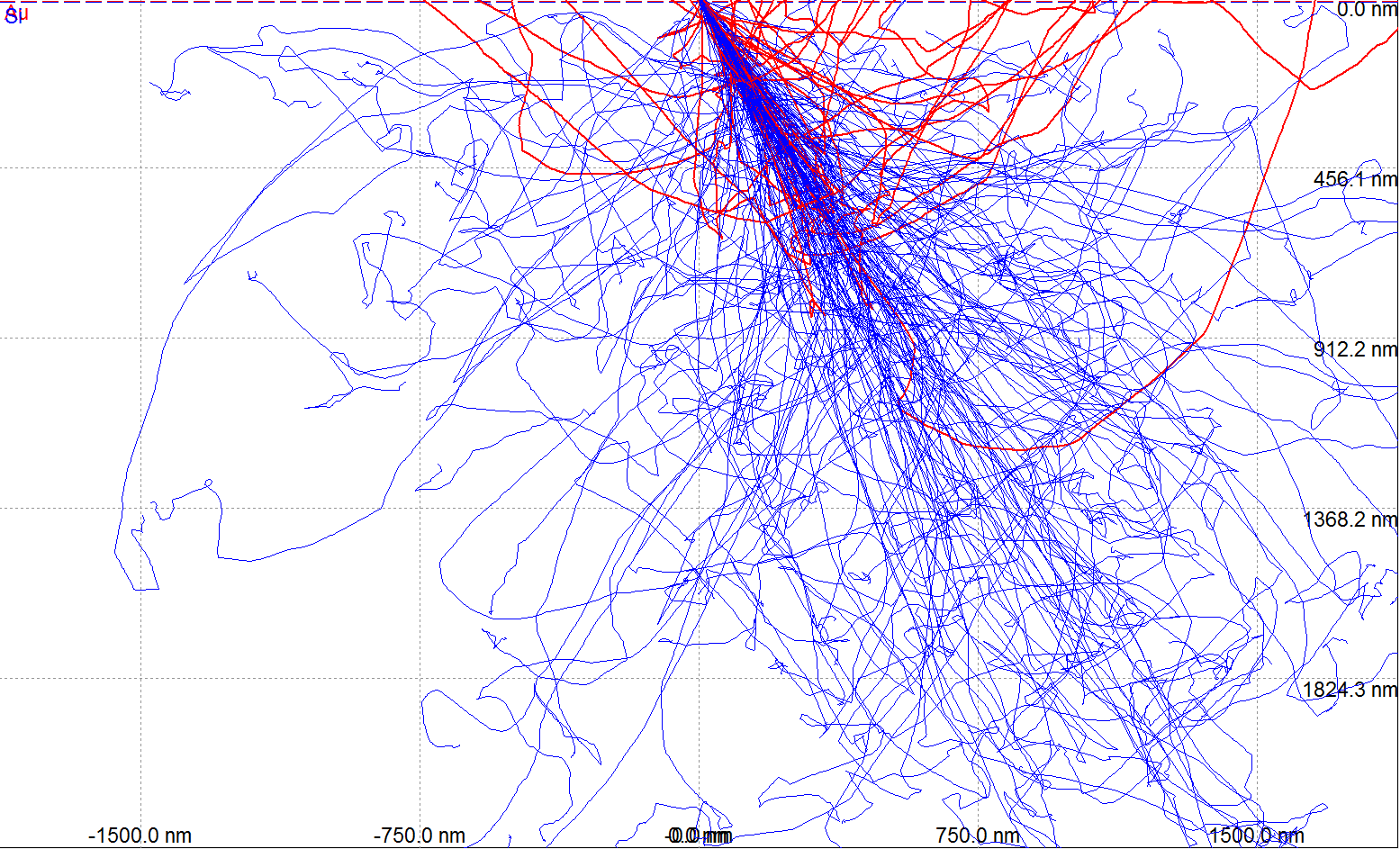}
\end{minipage}
{\caption{(Color online) On the left -- simulated electron tracks at energy of 18~keV falling onto the detector at 0 degree to the normal; on the right -- the same, but at 30 degrees. With red color back scattered electrons are shown .}
\label{fig:tracks}}
\end{figure}

Our spectrometer as a MAC-E filter is constructed in a such way that electron moving in variable magnetic field changes its angle relative to the spectrometer axis. This angle at the entrance and at the detector follows the relation $sin(\theta_{pinch})/sin(\theta_{detector})=\sqrt{B_{pinch}/B_{detector}}$. For field configuration of 7.2~T and 2.1~T in pinch and detector magnets, respectively, at the spectrometer entrance the angular range is from 0 to 90 degrees, while  at the detector position the range shrinks to 0 -- 33 degrees.  In Fig.~\ref{fig:tracks} we show how electrons move inside our detector at two impact angles as it was calculated by CASINO simulation. 
%%%%%%%%

\begin{figure}[htb]
\centering
	\includegraphics[width=0.7\linewidth]{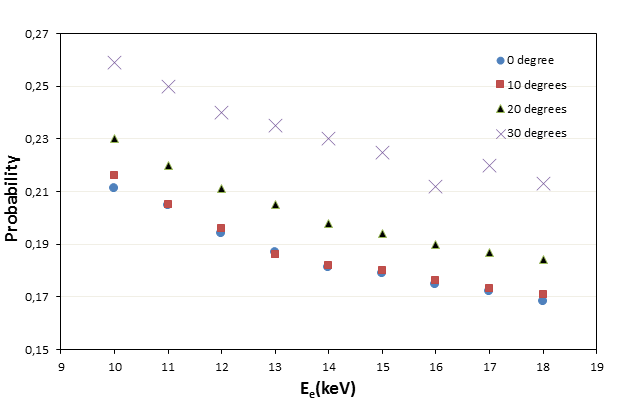} 
	\caption{Probability for backscattering versus electron energy for different impact angles.}
	\label{fig:prob}
\end{figure}

\begin{figure}[htb]
\centering
	\includegraphics[width=0.7\linewidth]{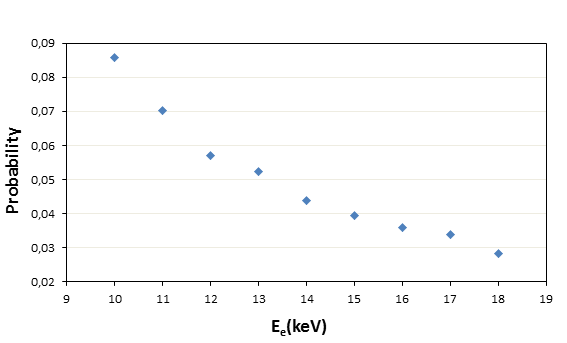} 
	\caption{Probability for backscattering on the 10 nm Au window versus electron energy at zero degree impact.}
	\label{fig:Au_scat}
\end{figure}

\begin{figure}[htb]
\begin{minipage}[t]{0.48\linewidth}
\includegraphics[width=1.\textwidth]{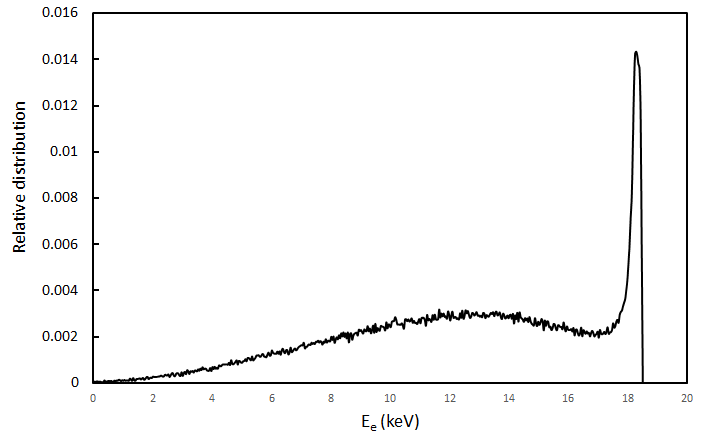}
\end{minipage}
\hfill
\begin{minipage}[t]{0.48\linewidth}
\includegraphics[width=1.03\textwidth]{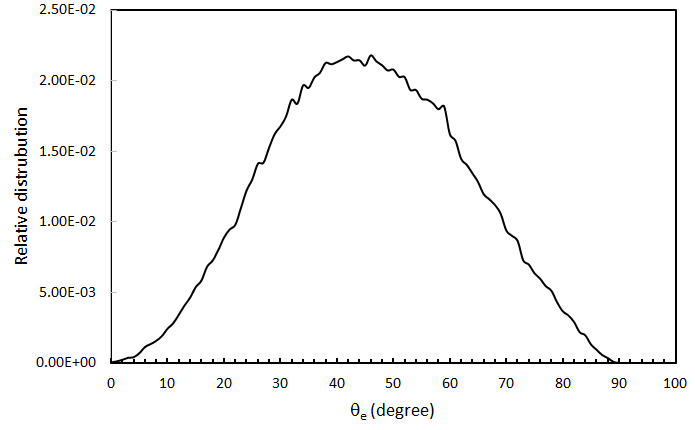}
\end{minipage}
{\caption{Incident electrons at18.5 keV and at 20 degrees. On the left -- energy spectrum of backscattered electrons.  On the right -- their angular distribution. Spectra are normalized to the total number of scattered electrons.}
\label{fig:E_sp}}
\end{figure}

There is a quite significant amount of tracks which scatter back from the detector. The probability  to escape depends on electron energy and impact angle, see Fig.~\ref{fig:prob}. 
The 10 nm gold window at  the entrance gives a smaller contribution to the scattering, at the level of a few percent, Fig.~\ref{fig:Au_scat}.
%%%%%%%%
%%%%%%%%
There are rather wide energy and angular distributions of scattered electrons, Fig.~\ref{fig:E_sp}. The energy spectrum has a specific peak on the right corresponding to scattering on the gold window. The wide energy distribution in the center corresponds to the electrons scattered from deep inside Si. The angular spectrum is rather independent of electron energy or impact angle. 
%%%%%%%%

\section{Destiny of backscattered electrons}
We can distinguish two classes of backward electron scattering: class~(A) -- scattering occurs on the Au window and do not give a signal in Si and class~(B) --  electron goes through the Au window,  scatters in Si and leaves some signal in Si. Each of these classes of events should also be split onto three groups: 
\begin{enumerate}
\item electrons with the energy lower than a high voltage (HV) potential on the spectrometer electrode; 
\item electrons with the energy higher than the applied HV which scatter back at angles larger than 33 degrees; 
\item the same as type 2, but at angles less than 33 degrees.
\end{enumerate}
The reason for a cut at 33 degrees is our MAC-E filter field configuration. The electron angle to the spectrometer axis in two different points 1 and 2 follows the  mentioned above, formula $sin\theta_1/sin\theta_2= \sqrt{B_1/B_2}$,  where $B$ is a magnetic field in these points. As a result, electron angle  of 90 degrees to the spectrometer axis in the pinch magnet translates to 33 degrees at the detector. This is also valid for scatter electrons moving in the opposite direction.  If their energy is above the spectrometer potential, and if they have angles larger than 33 degrees, they will be reflected by the pinch magnet. This magnet works as a magnetic mirror. Electrons at smaller angles will pass the pinch magnet and will get into the original electron source, gaseous tritium source in our case, and, as we assume now, will be lost there.

For our experiment the most crucial question  is how all these processes depend on electron energy and spectrometer setup. To answer this question an additional calculation of electron trajectory in the actual configuration of magnetic and electrostatic fields is needed.

\begin{figure}[htb]
\centering
	\includegraphics[width=0.4\linewidth]{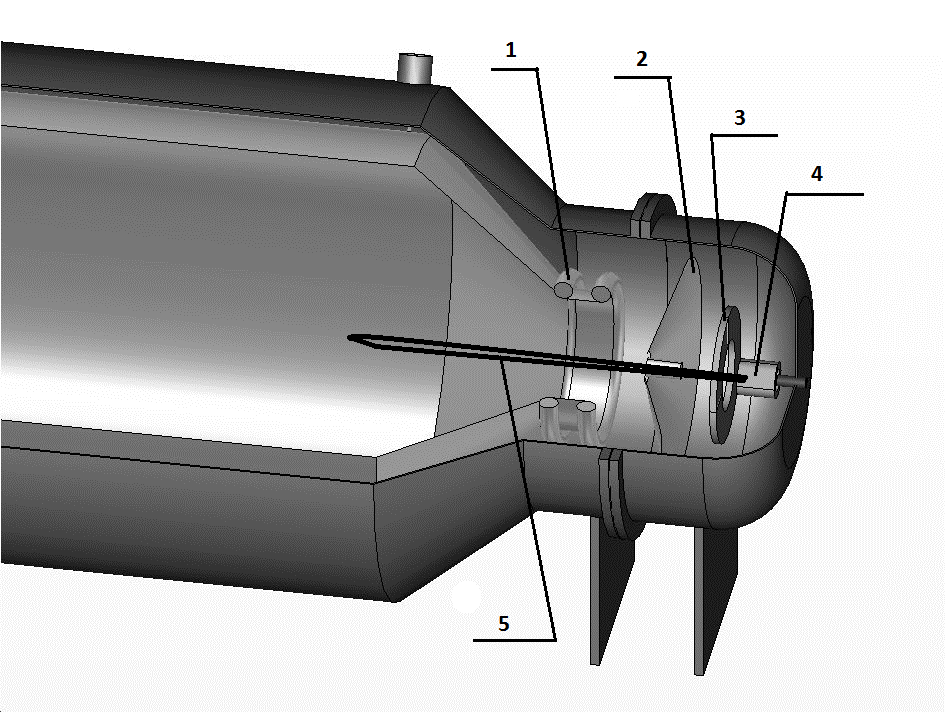} 
	\caption{Detector side of the spectrometer. 1 -- high voltage electrode, 2 -- ground electrode, 3 -- disk magnet, 4 -- 2.1~Tesla solenoid with detector inside, 5 -- track of scattered electron with energy less than electrode potential.}
	\label{fig:Det_Side}
\end{figure}

\begin{figure}[htb]
\centering
	\includegraphics[width=0.4\linewidth]{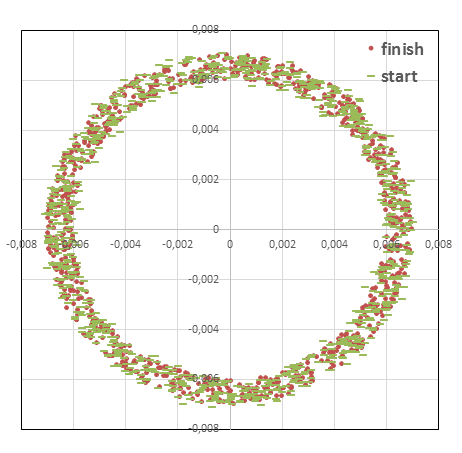} 
	\caption{Simulated X-Y position of started  electrons from a circle near the detector edge and reflected by electrostatic filed. Dimension are given in meters .}
	\label{fig:ring}
\end{figure}

Originally we started simulations by using ANSYS software, Release 14.0 © SAS IP, Inc. Then, to hasten calculation of magnetic and electrostatic fields in the spectrometer we use software packet of analytical calculations developed by F. Gl$\ddot{u}$ck~\cite{gluck1, gluck2} and used in computation of the KATRIN spectrometer. Electrons from group-\textbf{1} have lower energy and are electro-statically reflected by the spectrometer field. In Fig.~\ref{fig:Det_Side} we symbolically show such a track. \textbf{ALL} these tracks were  found to return to the detector almost at the same position, Fig.~\ref{fig:ring}. Electrons are not lost in this case, but there are two important factors. First, the returning electron will cover some distance and  comes with a delay relative to the time when the primary electron hits the detector.
%%%%%%%%%
%%%%%%%%

\begin{figure}[htb]
\centering
	\includegraphics[width=0.7\linewidth]{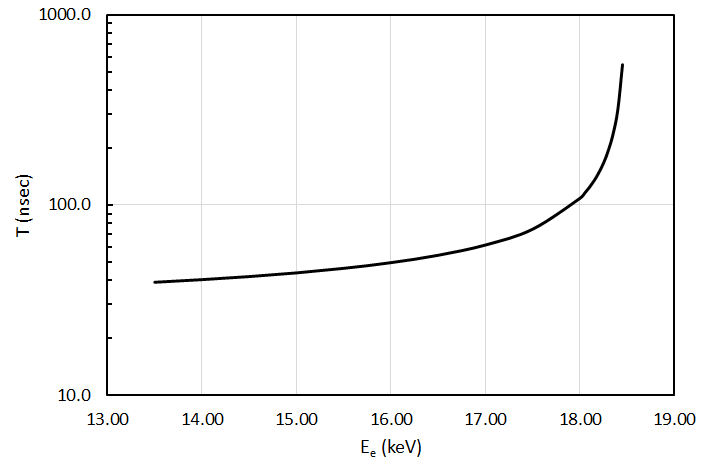} 
	\caption{Time delay between the moment when the primary electron hits the detector and time when scattered electron is reflected by the electrostatic mirror returns back to the detector versus the energy of scattered electron. The presented case is when spectrometer is at a 18500~V potential.}
	\label{fig:time}
\end{figure}

For example, in Fig.~\ref{fig:time} we present the case when the spectrometer is at a 18500~V potential and the primary electron energy is 18490 eV. The larger the energy of the scattered electron, the deeper may it fly into the spectrometer. For a distance of 1-1.5 meters from the detector the delay  reaches 500-600~nsec. The returning time depends on the scattered angle, but not much. 

Here arises a problem. If registration electronics is slow with a signal integration time of 1-2 $\mu$sec, as it was in the previous Troitsk nu-mass measurements, both signals are summed and registered as a single event, thus there is no double counting. If electronics is fast, we get two signals for the event from class-B: there is be some signal from the primary electron and later -- from the scattered and reflected electron. This is critical for the future electronics selection  choice. Second, another specific feature of the group-\textbf{1} events is that the signal amplitude in Si will be smaller at least by 1-1.5 keV because the scattered electron passes the non-sensitive Au layer twice.  
%%%%%%%%%

\begin{figure}[htb]
\centering
	\includegraphics[width=0.4\linewidth]{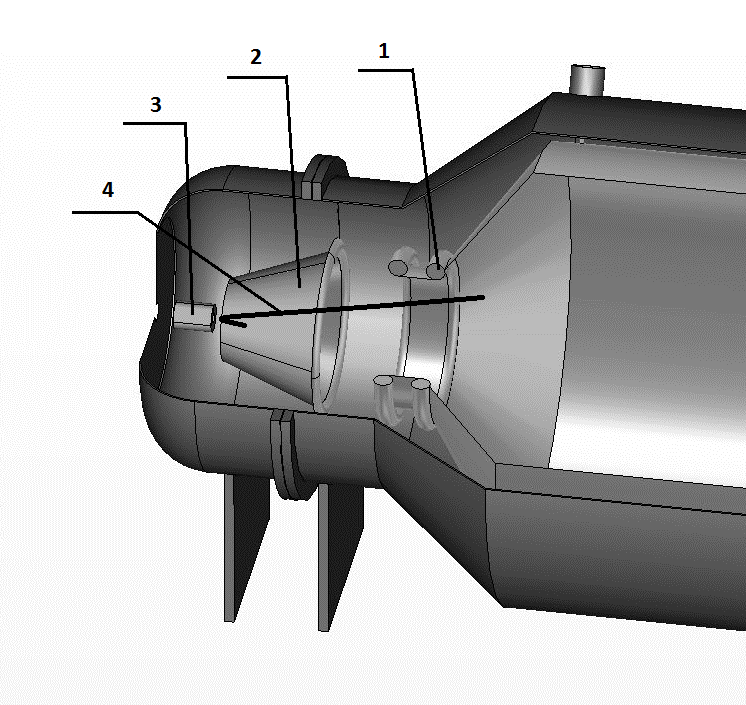} 
	\caption{Entrance of the spectrometer. 1 -- high voltage electrode, 2 -- ground electrode, 3 --  7.2~Tesla pinch magnet, 4 -- track reflected by magnetic mirror.}
	\label{fig:Entrance}
\end{figure}

As already mentioned, electrons from group-\textbf{2} and group-\textbf{3} pass electrostatic barrier and reach the pinch magnet, Fig.\ref{fig:Entrance}. The destiny of group-\textbf{3} is clear -- they will escape and will be lost. A question remains how many of them will be at each electrostatic potential when we measure tritium beta-spectrum in a wide energy range? 
%%%%%%%%%

\begin{figure}[htb]
\centering
	\includegraphics[width=0.5\linewidth]{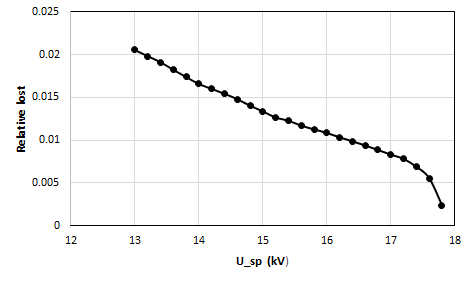} 
	\caption{The relative amount of electrons which will be lost by escaping the spectrometer through the left side versus the spectrometer potential. Primary electron energy is 18 keV.}
	\label{fig:gun18}
\end{figure}

\begin{figure}[htb]
\centering
	\includegraphics[width=0.7\linewidth]{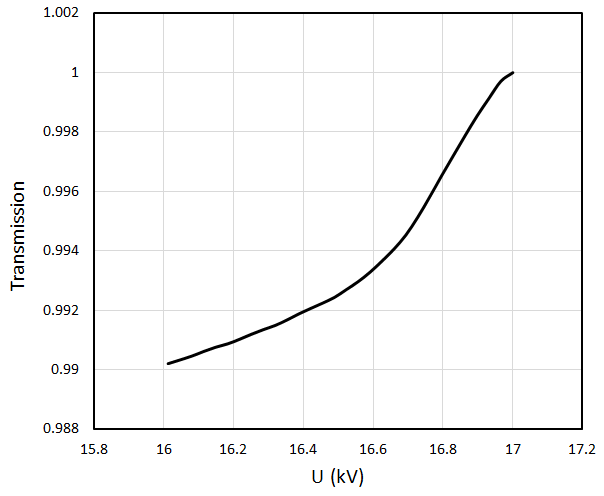} 
	\caption{The distortion of the spectrometer transmission function for  17~keV electrons versus potential $U$ on the spectrometer in 1 kV interval under the electron energy.}
	\label{fig:transm_f}
\end{figure}

First, let's find out how many electrons will escape at a fixed primary electron energy which hit the detector, say, from the calibration electron gun shooting at the spectrometer from the left side. The result of calculation is presented in Fig.~\ref{fig:gun18}. We take electron energy of 18 keV at an impact angle of 20 degrees and vary the spectrometer potential. To make an estimate in this particular case we use the probability to scatter from Fig.~\ref{fig:prob} (0.185), cut at angle less than 33 degrees (0.24, Fig.~\ref{fig:E_sp}, right) and then scan the scattered electron energy spectrum calculated by CASINO, Fig.~\ref{fig:E_sp}, left. The value of lost electrons is quite significant and reaches a few percent. There is some feature on the right at this plot. It reflects a peak in the energy distribution caused by scattering on the gold window, Fig.~\ref{fig:E_sp}. This effect distorts the spectrometer transmission function which supposed to be flat for the spectrometer potential smaller than the electron energy, Fig.~\ref{fig:transm_f}. The smaller the spectrometer potential $U$, the more scattered electrons are above this $U$ and  can escape from the spectrometer.

Such an effect is small but can distort the tritium beta-spectrum near the endpoint. Our estimations show that this effect is small  indeed compared to the magnetic trapping effect~\cite{Aseev:2011dq} in our windowless gaseous tritium source. Nevertheless, the distortion can change the measured slope of the tritium beta-spectrum at the very end point by shifting the extracted $m^2_{\nu}$ to negative region and should be taken into account. All these considerations are true for the planned experiment KATRIN to search for electron antineutrino mass limit~\cite{katrin_cdr} which has a very similar spectrometer.  
%%%%%%%%%

\begin{figure}[htb]
\centering
	\includegraphics[width=0.7\linewidth]{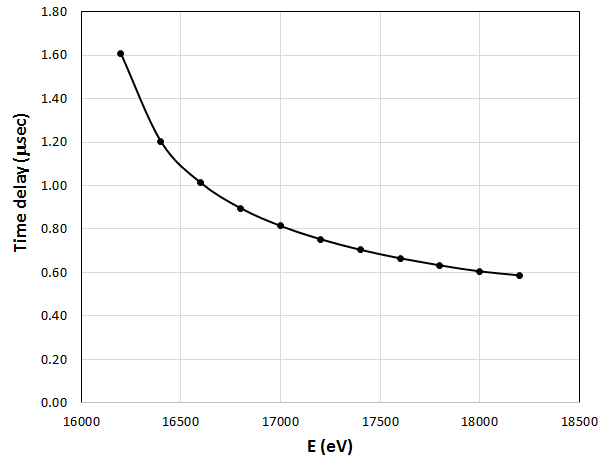} 
	\caption{Overall time for scattered electron to return to the detector after reflection by the magnetic mirror near the pinch magnet versus electron energy. The spectrometer potential is 16000~V, electrons are scattered at 50 degrees.}
	\label{fig:time_mirror}
\end{figure}

Another difficulty for correction is the group-\textbf{2} when the electron energy is higher than the applied HV and the electron scatters back at an angle larger than 33 degrees. All these electrons will be reflected by the magnetic mirror formed by the pinch magnet and pass the distance at least twice the size of the spectrometer. As a result, the returning time to the detector for such electrons can reach 1-2 microseconds, Fig.~\ref{fig:time_mirror}. 
%%%%%%%%%%

The good news is that these electrons should be energetic enough to pass the electrostatic barrier, thus they cannot loose much energy in the detector before scattering. In case of a fast registration electronics which can distinguish these two signals, the first one will be with a very low amplitude in Si detector and may be under the registration threshold. Thus, there is a very high probability that  the primary electron will not be counted twice and registered as a `'regular'' event. More careful estimation and correction can be done after the exact knowledge of the electronics response.  

\section{Effect on the sterile neutrino search}
\begin{figure}[htb]
	\centering
	\includegraphics[width=0.7\linewidth]{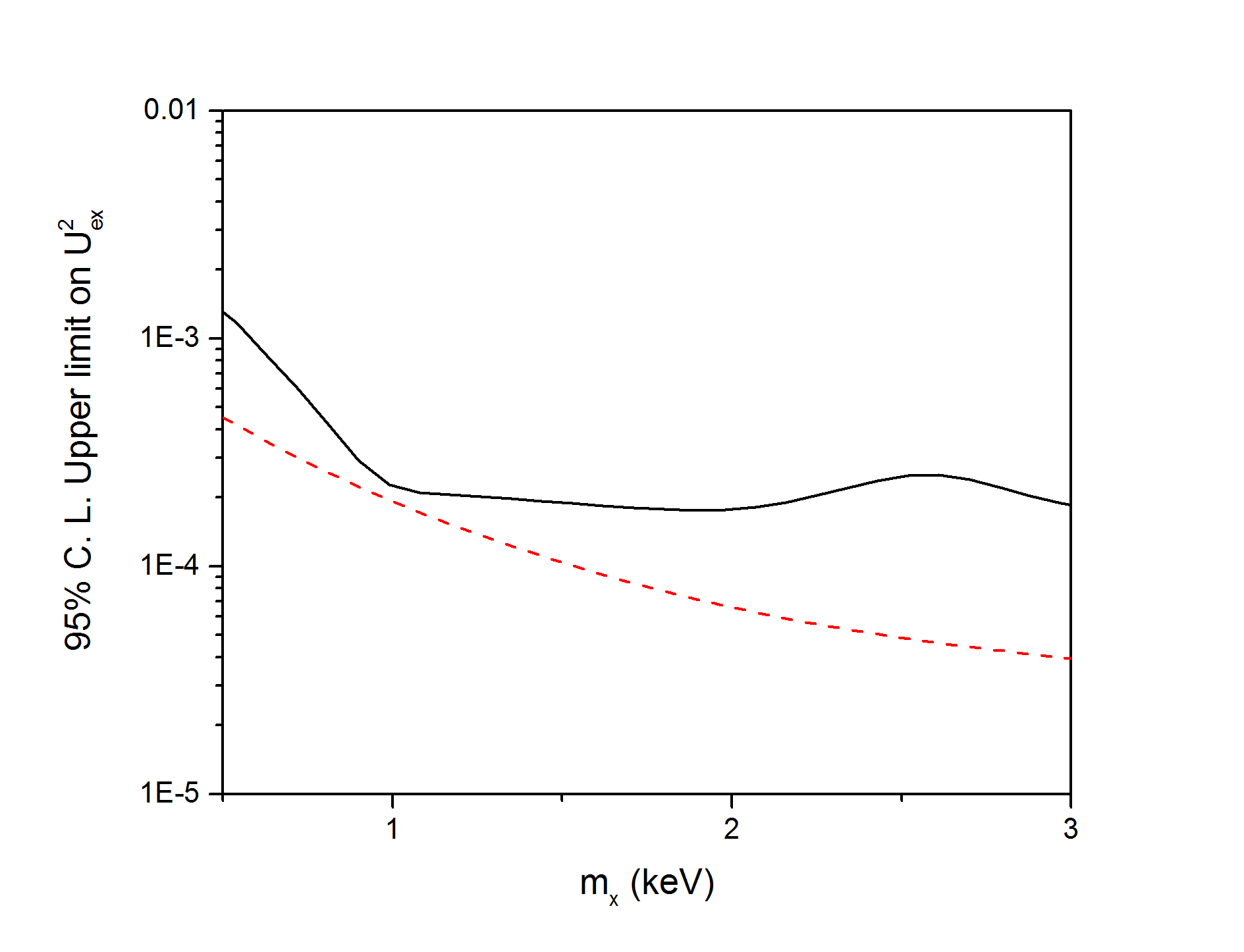} 
	\caption{The solid line depicts sensitivity for the sterile neutrino mixing matrix element $U_{ex}^2$ and the  dashed line reflects contribution  from the  1\% uncertainty of the effect of electron scattering as goals of the first stage of our measurement with existing detector and electronics~\cite{our_nus}. }
	\label{fig:limits}
\end{figure}

The uncertainty of a spectrometer and detector combined response function is one of the main sources of systematic errors for the planned sterile neutrino search. Fig.~\ref{fig:limits} shows  sensitivity limit (which is calculated assuming that this error is dominant) for sterile neutrino squared mixing matrix element $U_{ex}^2$ in case the relative uncertainty for the effect described in this article is 1\%. Such an error is lower than other expected systematic errors at the first phase of our experiment. Additionally we have found that if the electron trapping in our gaseous tritium source is set as a free parameter, these errors are greatly diminished. This comes from the fact that the trapping parameter will cover or contain the transmission function distortion induced by electron scattering,  see Fig.~\ref{fig:gun18}. The sum of these effects can be disentangled from the sterile neutrino kink by the analysis procedure.

\section{Conclusion}
We estimate the influence of electron scattering from a particle detector on the response of a MAC-E type spectrometer in the energy range up to 20 keV. For this, a code for electron interaction with solid materials, CASINO, and analytic calculation of magnetic and electrostatic fields were used. The value of scattered electrons reaches 20\%. Depending on electron scattered energy and angle, electron may be reflected by the electrostatic or magnetic mirrors and hit again the detector but with a delay which can reach a few hundred nanoseconds for the electrostatic reflection and two microseconds for the magnetic mirror. There is also a case when electron may run away through the spectrometer entrance. Understanding and careful estimation of all such effects is critical for the planned experiment  on search for a sterile neutrino in a few keV mass range. It should also be taken into account for making decision on future electronics design. The probability for electrons to escape from the spectrometer distorts the spectrometer transition function, especially for tritium beta decay measurements at the spectrum endpoint. The ignorance of correction for such an effect may also move the value of the extracted electron antineutrino mass, $m^2_{\nu}$. The exact estimation requires the knowledge of detector and registration electronics details.

\section{Acknowledgments}
We thank Ferench  Gl$\ddot{u}$ck who kindly gave us permission to use his code and Victor Matushko for useful discussions. This work was partially supported by RFBR under grant numbers 14-02-00570-a and 14-22-03069-ofi-m.
%\bibliographystyle{JHEP}
%\bibliography{Det_Scatt}

\begin{thebibliography}{20}
\bibitem{Aseev:2011dq}V.N.~Aseev et al., Phys.Rev., \textbf{D84} (2011) 112003. 
\bibitem{our_nus}D.N.~Abdurashitov et al., JINST \textbf{10} (2015) 10, T10005; arXiv:1504.00544. 
\bibitem{katrin}S.~Mertens et al., arXiv:1409.0920.
\bibitem{ptolemy}S.~Betts et al., arXiv:1307.4738.
\bibitem{casino} http://www.gel.usherbrooke.ca/casino/index.html.
\bibitem{gluck1} F. Gl$\ddot{u}$ck, Progr. Electromagn. Res., \textbf{32}, (2011) 319.
\bibitem{gluck2} F. Gl$\ddot{u}$ck, Progr. Electromagn. Res., \textbf{32}, (2011) 351.
\bibitem{katrin_cdr} KATRIN CDR, https://www.katrin.kit.edu/publikationen/DesignReport2004-12Jan2005.pdf.
\end{thebibliography}

\end{document}